
%
%
\magnification=1200

\def\today{\ifcase\month\or
  January\or February\or March\or April\or May\or June\or
  July\or August\or September\or October\or November\or December\fi
  \space\number\day, \number\year}
\font\twelverm=cmr12  \font\twelvebf=cmbx12
\font\tenrm=cmr10

\font\textfont=cmr10 at 12.truept

\def\bigtype{\let\rm=\twelverm \let\bf =\twelvebf \rm}

\def\hsingle{\baselineskip=20truept}

\overfullrule=0pt
\vsize=8.8truein
\hoffset=.2truein
\hsize=6.1truein
\hsingle

\parskip=5pt plus 2pt
\parindent=20truept

\tenrm

\vskip 1cm
\centerline{\bigtype{\bf
Interface depinning in a disordered medium - numerical results
}}

\bigskip\bigskip
\textfont
\centerline{Heiko Leschhorn}
\bigskip
\centerline{Theoretische Physik III, Ruhr-Universit\"at Bochum,}
\smallskip
\centerline{Postfach 102148, D-4630 Bochum, Germany}

\bigskip
\vskip .5truecm
{\narrower\smallskip\noindent
{\bf Abstract:}
We propose a lattice model to study the dynamics of a driven interface
in a medium with random pinning forces. The critical exponents
characterizing the depinning transition are determined numerically
in 1+1 and 2+1 dimensions. Our findings are compared with recent numerical
and analytical results
for a Langevin equation with quenched noise,
which is expected to be in the same universality class.

\smallskip}
\bigskip

\vfil\eject

\tenrm

{\bf I. Introduction}

In recent years there has been considerable interest in the scaling
behaviour of interfaces and directed lines
broadened by randomness [1] and
in the critical phenomena which occur at the onset of steady-state
motion [2,3].
The driven viscous motion of an interface in a medium
with random pinning forces combines aspects of both of these two fields.
An example is the motion of a
domain wall in the random field Ising model [4,5].
The relaxation of metastable domains in a diluted
antiferromagnet with applied external magnetic field
provides a possible experimental realization [6]. Due to random field pinning
the domain walls become rough and if the driving field is sufficiently
weak compared to the random field strength, the domain walls
become stuck and the domain state is frozen.
Random field pinning can also be important for fluid displacement
experiments in porous media [7], where an anomalous roughness in comparison
with the KPZ-equation [1] was observed.
Kessler, Levine and Tu [8] (KLT) and recently
He, Kahanda and Wong [7] suggested that the increased roughness
is due to {\it quenched} random capillary forces on the interface.
(Another interesting proposal [9] is to assume that the noise
fluctuates in time but follows a power-law distribution. This possibility
will not be considered here.)
Very similar problems arise in vortex pinning
in type-II superconductors [10] and in sliding charge-density waves [2,11].

The simplest continuum description for the driven motion of an interface
in a random medium is given by the following Langevin equation for
a $D$-dimensional interface profile $ z({\bf x},t) $ [12,13]
$$ \lambda {\partial z \over \partial t} = \gamma
\nabla ^2 z + F + \eta ({\bf x} , z) \eqno(1) $$
where $\lambda $ and $ \gamma $ are the inverse mobility and
the stiffness constant, respectively, and $F$ is a uniform driving force.
The random force $\eta ({\bf x} , z) $ is Gaussian distributed with
$\langle \eta \rangle = 0$ and correlations
$\langle \eta ({\bf x_0},z_0) \eta ({\bf x_0}+{\bf x},z_0 +z) \rangle
= \delta ^ D ({\bf x}) \Delta (z)$.
For the random field case the correlator $\Delta (z)$ is a monotonically
decreasing function of $z$ for $z>0$ and decays rapidly to zero over a finite
distance.
The difference to the Edwards-Wilkinson (EW) equation [14] is that the noise
depends on the position of the interface, which makes the problem
highly nonlinear. However, the noise that acts on a sufficiently
fast moving interface is fluctuating in time
like in the EW equation. Therefore a crossover
to a EW regime is expected as the velocity increases to large values.
Here we are interested in the critical behaviour at the onset of motion
where the external driving force is just able to overcome the pinning
forces. The velocity $v$ of the interface scales with the reduced
force $f=(F-F_c)/F_c$ as $v \sim f^\theta $ where $F_c$ is the
threshold force [2,15].
It can be shown [15] that there is a correlation length $\xi$ which
diverges with $F  \to F_c$ as $\xi \sim f ^{-\nu }$.
Thus, at $F=F_c$, the roughness (or interface width)
$$w^2 (L,t) = \Bigl \langle \overline {[z({\bf x},t) - \overline
{ z( {\bf x'},t)}]^2 } \Bigr \rangle \eqno (2) $$
scales as [16]
$$ w (L,t) \sim L^\zeta \Psi \left ( {t \over L^z } \right ) \eqno (3) $$
where $L$ is the system size, $\zeta$ is the roughness exponent,
$ z $ is the dynamical exponent and $\Psi$ is a scaling function with
$ \Psi (y) \sim y^\beta $ ($\beta \equiv \zeta/z$) for $y\ll 1$
but becomes constant for $ y\gg 1 $.
In eq.(2) the overbar denotes the spatial average over ${\bf x}$
and the angular brackets mean the configurational average.
It is known [12,15] that above four interface dimensions $D$
the interface is flat, i.e. $\zeta = 0$. For $D=4- \epsilon ,
 ~ \epsilon \ll 1 $ the
critical exponents were calculated by a functional renormalization
group scheme to first order in $\epsilon $: $\zeta \simeq \epsilon /3$,
$~z=\zeta / \beta \simeq 2- 2 \epsilon /9 $, $~\nu \simeq 2/(6-\epsilon) $
and $ ~\theta = \nu (z-\zeta) $ [15].
The equation of motion (1) was simulated by KLT in 1+1 dimensions
for a wide range of velocities [8].
For $v \to 0 $ they suggest that $ \zeta \to 1 $
which would be in agreement with an Imry-Ma argument [4].
Parisi [17] proposed that the argument $z$ of
the random force $\eta$ in eq.(1) can be replaced by a constant from which
follows $\zeta = (4-D)/2$, $\beta =(4-D)/4$ and $\theta =1$ for $D \le 4 $.
He numerically determined $\beta$ in $D=1,2$ and $3$ from (1) with results
consistent with his analytical arguments.
Recently, Tang [18] used the Runge-Kutta scheme to solve eq.(1)
and found in 1+1 dimensions $\zeta= 1.25 \pm 0.10$, $\beta = 0.81 \pm
0.02 $, $ ~\theta = 0.4 \pm 0.05 $ and $\nu = 1.1 \pm 0.1$,
in disagreement with both KLT [8] and Parisi [17].

Here we introduce a lattice model of probabilistic cellular automata [19]
which is expected to be in the same
universality class as eq.(1). In 1+1 dimensions the results are
$\zeta (D=1) = 1.25 \pm 0.01, ~\beta (D=1) = 0.88 \pm 0.02, $ and in
2+1 dimensions $\zeta (D=2) = 0.75 \pm 0.02, ~\beta (D=2) =0.475 \pm 0.015$
and $\theta (D=2) = 0.65 \pm 0.05 $. These values support
the results of the $\epsilon $-expansion [15] because the deviations for
$4-D = \epsilon = 2$ are much smaller than for $\epsilon = 3$.

\bigskip
{\bf II. The model}

In order to simplify the notation the model is explained for the case of
1+1 dimensions. The generalization to higher dimensions is straightforward.
Consider a square lattice where each cell $(i,h)$ is assigned a random
pinning force $ \eta _{i,h}$ which takes the value 1 with probability $p$
and -1 with probability $q=1-p$.
By excluding overhangs the interface
is specified by a set of integer column heights $h_i (t) $, $i=1,2...L$.
At $t=0$ all columns have the same height $h_i (t=0) = 0 $. During the
motion for a given time $t$ the value
$$ v_i = h_{i+1} (t) + h_{i-1} (t) - 2 h_{i} (t) + g \eta _{i,h} \eqno (4) $$
is determined for all $i$, where $g$ is a parameter, $\eta_{i,h} =\pm 1$
and periodic boundary conditions are used.
The interface configuration is then updated  simultaneously for all $i$:
$$ \eqalignno{h_i (t+1) &= h_i (t) +1 ~~~~~~~ if ~~v_i > 0 & \cr
h_i (t+1) &= h_i (t) ~~~~~~~~~~~~otherwise. & (5) \cr }$$
The parameter $g$ measures the strength of the random force compared to the
elastic force and
the difference $p-q$
determines the driving force.
The growth rule specified by eqs.(4) and (5)
can be motivated by the continuum equation (1)
because $v_i$ is the sum of the discretized Laplacian and a random
force. It remains however an open question whether the use of a
non-Gaussian noise and the discretization of $x,z$ and $t$ changes
the universality class on accessible length scales.
The relation of this model to the mechanism introduced earlier [20]
for pinning by directed percolation will be discussed below.

\bigskip
{\bf III. Numerical results}

\noindent a) 1+1 dimensions

Since we are interested in the scaling behaviour at the depinning transition
one has to find first
the critical value $p_c (g) $ of the concentration of plus cells where the
infinite system gets pinned.
For $g=1$ we estimate $p_c (g=1) \simeq 0.8004 $,
while  $p_c (g=2) \simeq 0.8748$.
To determine the roughness
exponent $\zeta$, different system sizes with $ 8 \le L \le 1024 $ were
simulated at $p=p_c$ until they get pinned. The values for $w^2 (L) $
were averaged over several thousand independent runs $n$, depending on $L$
($ ~n \sqrt L \approx 8*10^4 $).
Fig.1 shows $w^2 (L) $ in a double logarithmic plot. For
$ 32 \le L \le 512 $
the data are well fitted by a straight line from which we get
$\zeta (D=1) = 1.25 \pm 0.01 $. Error bars indicate only statistical
uncertainities.
The fluctuations are somewhat larger for $g=2$ than for $g=1$.
For bigger $L$ the simulations become very time consuming because
it takes longer for a larger
system to get pinned at $p=p_c$. In addition, when one increases $L$,
one has to know the threshold $p_c$ more accurately to assure $\xi \gg L$.
To see the deviations from the scaling behaviour we define an
effective roughness exponent
$\zeta (L) \equiv \log[w(2L)/w(L)]/\log2 $. In fig.2 $~\zeta (L)$ for
$g=1$ and $p=p_c$ is compared with $\zeta (L)$ for $g=2$ and $p=0.8744 < p_c$.
For $L=256$ and $512$, $~\zeta (L) $ is significantly smaller for $p<p_c$ than
for $p=p_c$.

Besides the possibility to determine the dynamical roughness exponent $\beta$
from the width $w(t)$ it is also possible to consider the scaling behaviour of
$H(t) \equiv \langle \overline{h_i} \rangle \sim t^\beta $.
{}From the latter we get $\beta (D=1) = 0.86 \pm 0.02 $ for $L=16384$,
while from the width $w(t)$ $\beta (D=1) = 0.87 \pm 0.04 $.
The scaling behaviour of $H(t)$ for larger systems with $L=262144$
is plotted in fig.3,
while in fig.4 the width $w^2 (t)$ is shown for the same runs. The data
were averaged over 40 independent runs. (This took 21.5 h CPU on one
Cray Y-MP processor. The other data presented in this paper required
about 2000 h CPU on a IBM RS/6000 model 320 workstation.)
The effective exponents
$\beta (t) \equiv \log[w(2t)/w(t)]/\log2 $ (and analogously for $H(t))$
are shown in fig.5a.
A possibility to take into account corrections to scaling is to plot
$w^2 (2t) - w^2 (t) $ versus $t$.  The corresponding effective exponents
are shown in fig. 5b. The exponents determined by the width remain
unchanged.
Neither is there a clear improvement for the scaling of the
height, although the plateau of $\beta (t)$
seems to be larger but also the fluctuations are bigger in fig. 5b.
Note that the uncertainities in the last two points are much larger,
so we do not expect that $\beta (t)$ increases
to larger values.
Thus, we conclude $ \beta (D=1) = 0.88 \pm 0.02 $.
We have checked that this value is consistent with the case $g=2$.
The height-height correlation function
$ C^2(r,t)=\langle \overline {[h_{i+r}(t) -  h_i (t)]^2}   \rangle$
is expected to scale in the same way as the width in eq.(3).
Fig.6 shows a scaling plot, where $C(r,t)$ is divided by $t^\beta$
and $r$ by $t^{1/z}$. The best data collapse is achieved for
$\zeta \simeq 1.25 $ and $\beta \simeq 0.88 $, in complete agreement
with the above results.

For large $p=0.95$ and $0.97$ the velocity is about $v=0.78$ and
$0.88$, respectively ($g=2$).
{}From the interface width $w(t)$ we find
$ \beta (p \gg p_c)= 0.25 \pm 0.01 $, which is in agreement with the expected
EW [14] regime far away from the depinning transition.
We were not able to estimate the exponent $\theta$. Due to the large
roughness the fluctuations of $v$ become too large to fix the
exponent $\theta $ with a reasonable accuracy.

\noindent b) 2+1 dimensions

As in 1+1 dimensions we determine $\zeta $ from the simulation of
interfaces with different sizes until they get pinned. Fig.7 shows
the effective exponents $\zeta (L) $ for $g=4,~p=p_c\simeq 0.6416$
and $g=6,~p=p_c \simeq 0.74448$ from which we estimate
$\zeta = 0.75 \pm 0.02 $.
The scaling behaviour of the interface width for a system
of size $L^2 = 1024^2$
is shown in fig.8 from which follows $\beta (D=2) = 0.475 \pm 0.015 $.
(The data were averaged over 15 independent runs.)
The scaling of $H(t)$ has to be considered
with caution as will become clear from fig.9 where the effective
exponents $\beta (t) $ are plotted.
For $t \gg t_{\times} = A L^z $, $~H(t)$
goes to a constant for $p<p_c$ or grows linearly for $p>p_c$,
while the scaling $H(t) \sim t^\beta $ is expected for
$ t_a \ll t \ll t_{\times} $ where
$t_a $ is the time where $\beta (t)$ achieves its asymptotic
value. The plateau of $\beta (t)$ determined by $H(t)$ for $p>p_c$
in fig.9 (triangles) for $32 \le t \le 128$ may be attributed to
a very slow crossover
because $t_a $ is of the same order as $ t_{\times}$.
Also for $p=p_c$, $\beta (t) $ (circles) never
achieves its asymptotic value. Thus, also the value for $\beta(t)$
determined from the scaling of the interface height $H(t)$ in 1+1 dimensions
has to be considered with caution.
For $g=6$ we find the same value $\beta = 0.475 \pm 0.015 $.
(For $g=1$ the velocity is nonzero even for $p<q$ which would
correspond to "negative driving forces". This strange situation
is perhaps due to the fact that the growth rule (5) does not allow $h_i$ to
decrease. Thus we choose $g$ such that $p_c>1/2$.)
The best scaling plots of $C(r,t)$ are achieved for
the parameters $\zeta = 0.74 $ and $\beta = 0.47...0.48 $ consistent with
the values given above.

In fig. 10 the velocity is plotted versus $p-p_c$ and with
$v \sim (p-p_c)^\theta $ we get $\theta = 0.65 \pm 0.05 $.
Although the fluctuations of $v$ are smaller in $D=2$ than in $D=1$
the uncertainties remain rather big. For $g=4$ and very small $v$,
$\theta$ seems to be about 0.68 but it then crosses over to
$\theta =0.64$. The exponent $\theta$ is very sensitive to the value of
$p_c$ for small $v$. For larger $v$, on the other hand, it is not clear
whether the scaling law $v \sim (p-p_c)^\theta $ still holds. It is known
from the charge-density waves that the critical region is quite narrow [2].
So the error bar on $\theta$ is
larger than those on the roughness exponents $\zeta $ and $\beta$.

\bigskip
{\bf IV. Discussion}

{}From the two roughness exponents
$ \zeta $ and $\beta $ we can calculate the dynamical exponent
$ z \equiv \zeta / \beta $: $~z(D=1) = 1.42 \pm 0.03$ and
$z(D=2) = 1.58 \pm 0.04 $,
i.e. the dynamics at the depinning transition is superdiffusive.
The correlation length exponent $\nu $ can be found from the general scaling
relation $ \nu = \theta / (z - \zeta) $ [15]. In 2+1 dimensions we get
$ \nu (D=2) = 0.8 \pm 0.05 $. Simple arguments suggest [15] that there
is a Harris criterion [21] for a sharp 2nd order depinning transition:
$1/\nu \le (D+\zeta)/2 $. In our case this relation is indeed fulfilled as
an inequality.

The simulations of the continuum equation (1) by Tang [18] yield a
dynamical roughness exponent
$\beta (D=1) = 0.81 \pm 0.02 $ which was determined from the scaling
of the height $H(t)$. This value is somewhat smaller than our
result $\beta = 0.88 \pm 0.02$.
However, we have seen that it is rather difficult to settle $ \beta $;
the fluctuations of the width are large and the height shows slow
crossover phenomena, especially for smaller systems.
Another possibility is that due to the special discretization
the dynamics of the lattice model is not described by the continuum
equation (1), e.g.
according to the growth rule eq.(5) there are only two velocities,
zero and one, whereas in the continuum model the velocity depends
continuously on the force.
On the other hand
the static roughness exponent $\zeta \simeq 1.25 $
agrees with the simulation [18] of eq.(1)
but not with the Imry-Ma result [8] ($\zeta = (4-D)/3 $ for all
$D \le 4$)
nor with the conjecture of Parisi [17] that $\zeta = (4-D)/2,
 ~ \beta = (4-D)/4,~\theta =1 $.
Our numerical values for the critical exponents support
the perturbative renormalization group expansion in
$ D=4-\epsilon,~\epsilon \ll 1 $ [15],
because the extrapolation to $\epsilon > 1 $ gives a better estimate
for $\epsilon =2$ than for $\epsilon = 3$.
It is interesting to note that the scaling relations
$z=\zeta / \beta = 2 - (\epsilon-\zeta)/3 $ and
$ \theta = 1 - {1 \over 3} (\epsilon-\zeta)/(2-\zeta) $
which were derived for small $\epsilon $ [15] are perfectly fulfilled by our
numerical results, although we have no arguments why they
should be exact.
By inserting our $\zeta$ in the scaling relations for $D=1$ we obtain
$ \theta (D=1) \simeq 0.22 $ and $ \nu (D=1) \simeq 1.33 $.

Our roughness exponents are much larger than those of the EW model and
the KPZ equation [22], i.e. quenched noise roughens an interface more than
thermal noise. This difference corresponds to the observed crossover
from the behaviour at the threshold ($\beta (D=1) \simeq 0.88$) to
that of large velocities ($ \beta (D=1) \simeq 0.25 $) where the noise
is short-range correlated in time.
Next, we discuss the relation to the growth mechanism
"pinning by directed percolation" [20],
where the dominating noise is also quenched. There, if the height difference
between two neighbouring columns exceeds a certain small value, the lower
column grows, independent of random pinning forces. Thus, the difference
between neighbouring columns is kept small and the growth is side-ways.
It can be shown [18] that the model is in the same universality class
as eq.(1) if a KPZ-term
$F ( \nabla z )^2 / 2 $
is added, which favours lateral growth. In the model presented here,
however, only the discretized Laplacian, i.e. the {\it sum} of the
height difference to {\it all} nearest neighbours enters the growth
rule (5). Therefore, the motion is not side-ways, which corresponds
to the absence of the gradient-squared term $F ( \nabla z )^2 / 2 $
in eq.(1), and arbitrary large slopes are possible which can occur
when the height difference to two neighbours has opposite sign.
Indeed, our roughness exponents are larger than those of an interface
pinned by a directed percolation cluster
($\zeta (D=1) = \beta (D=1) \simeq 0.63$
at the threshold [20]).
In 1+1 dimensions the exponent $\zeta \simeq 1.25 > 1$ is even unphysical.
If one derives eq.(1) from a Hamiltonian [1,4] the gradient terms
have to be assumed
to be small. If however $\zeta \ge 1$ the neglection of overhangs and
higher order gradients is no longer justified.
It is therefore questionable to apply this simple model to real 1+1
dimensional systems. For the equilibrium case of the random field
Ising model it is known [4] that $d=D+1=2$ is the lower critical
dimension $d_l$. At and below $d_l$ no long range order can be established
and the domain walls are convoluted in contradiction with the assumption
of no overhangs.

In 2+1 dimensions a recent simulation [5] of the motion of a domain wall
in the random field Ising model yields a roughness exponent
$\zeta = 0.67 \pm 0.03 $ which was obtained by a plot of the height-height
correlation function $C(r)$ for systems up to linear size 300.
In our measurements $C(r)$ never reaches its asymtotic scaling with
$\zeta \simeq 0.75 $ due to finite time and finite size effects
but gives effective exponents about 2/3. It would be therefore
very interesting to measure the width of a domain wall for different
system sizes to check whether the domain walls in the random field
Ising model of ref.[5] are correctly described by the model
presented in this paper.

\bigskip
{\bf V. Summary}

We have proposed a lattice model for the driven motion of an interface
in a random medium. The critical exponents characterizing the depinning
transition are determined in one and two interface dimensions.
We believe that the model is in the same universality
class as the continuum equation (1).
Our results (e.g. $\zeta (D=1) \simeq 1.25 , ~\zeta (D=2) \simeq 0.75 $)
are inconsistent with the validity of an Imry-Ma argument [8]
($\zeta = (4-D)/3, ~D \le 4 $) and with the approximation in eq.(1)
to neglect the $z$-dependence of the random force $\eta$ [17]
($\zeta = (4-D)/2$), but they support the analytical results of the
renormalization group for $D=4-\epsilon,~\epsilon \ll 1$ [15]
($\zeta = \epsilon /3$), because
the deviations are smaller for $\epsilon =2$ than for $\epsilon =3$.
It is of course desirable to investigate whether this difference
decreases further when $\epsilon = 1$.

\bigskip
\bigskip
{\bf Acknowledgements}

I am indebted to T.Nattermann, D.Stauffer and L.-H.Tang for useful
discussions and helpful comments on the manuscript.
I thank the supercomputer center HLRZ/J\"ulich for providing me
with computer time and H.-O. Heuer for his help using the Cray Y-MP
and for interesting discussions.
This work was supported by the Deutsche Forschungsgemeinschaft
through SFB 166.

\vfil\eject

{\bf References}
\par
\frenchspacing

\item{[1]} M. Kardar, G. Parisi, and Y.-C. Zhang,
Phys. Rev. Lett. {\bf 56} (1986) 889;
for a reviev see e.g.
J. Krug and H. Spohn, {\it Solids Far From Equilibrium: Growth, Morphology,
and Defects,} edited by C.Godriche (Cambridge University Press,
Cambridge, 1990)

\item {[2]} D. S. Fisher, Phys. Rev. Lett. {\bf 50} (1983) 1486;
O. Narayan and D. S. Fisher, Phys. Rev. Lett. {\bf 68} (1992) 3615.

\item{[3]} P. Bak, C. Tang, and K. Wiesenfeld, Phys. Rev. Lett. {\bf 59} (1987)
381; T. Hwa and M. Kardar, Phys. Rev. A {\bf 45} (1992) 7002.

\item{[4]} For recent reviews on the closely related equilibrium problem
see, e.g., T. Nattermann and P. Rujan, Int. J. Mod. Phys. B {\bf 3}
(1989) 1597; G. Forgacs, R. Lipowsky, and Th. M. Nieuwenhuizen,
in {\it Phase transitions and critical phenomena}, Vol. 14, edited by C. Domb
and J. L. Lebowitz (Academic Press, London, 1991), p.135.

\item{[5]} H. Ji and M. O. Robbins, Phys. Rev. B {\bf 46} (1992) 14519.

\item{[6]} S. Fishman and A. Aharony, J.Phys.C {\bf 12} (1979) L729;
S. J. Han, D. P. Belanger, W. Kleemann and U. Nowak,
Phys.Rev.B {\bf 45} (1992) 9728;
M. Lederman, J. Selinger, R. Bruinsma, J. Hammann and R. Orbach,
Phys.Rev.Lett. {\bf 68} (1992) 2086.

\item{[7]} M. A. Rubio, C. A. Edwards, A. Dougherty, and J. P. Gollub,
Phys. Rev. Lett. {\bf 63} (1989) 1685; S.-j. He, G. Kahanda, and
P.-z. Wong, Phys. Rev. Lett {\bf 69} (1992) 3731.

\item{[8]} D. A. Kessler, H. Levine, and Y. Tu,
Phys. Rev. A {\bf 43} (1991) 4551.

\item{[9]} Y.-C. Zhang, J.Phys. France {\bf 51} (1990) 2129;
Physica A {\bf 170} (1990) 1.

\item{[10]} A. I. Larkin and Yu. N. Ovchinikov, J. Low Temp. Phys. {\bf 34}
(1979) 409.

\item{[11]} K. B. Efetov and A. I. Larkin, Sov. Phys. JETP {\bf 45} (1977)
1236.

\item{[12]} R. Bruinsma and G. Aeppli, Phys. Rev. Lett. {\bf 52} (1984) 1547;
M. V. Feigel'man, Sov. Phys. JETP {\bf 58} (1983) 1076.

\item{[13]} J. Koplik and H. Levine, Phys. Rev. B {\bf 32} (1985) 280.

\item{[14]} S. F. Edwards and D. R. Wilkinson, Proc. R. Soc. London,
Ser. A {\bf 381} (1982) 17.

\item {[15]} T. Nattermann, S. Stepanow, L.-H. Tang, and H. Leschhorn,
J. Phys. II France {\bf 2} (1992) 1483.

\item{[16]} F. Family and T. Vicsek, J.Phys.A 18 (1985) L75.

\item{[17]} G. Parisi, Europhys. Lett. {\bf 17} (1992) 673.

\item{[18]} L.-H. Tang, unpublished.

\item{[19]} S. Wolfram, {\it Theory and Applications of Cellular Automata},
World Scientific, Singapore 1986.

\item{[20]} L.-H. Tang and H. Leschhorn, Phys. Rev. A {\bf 45} (1992) R8309;
S. V. Buldyrev, A.-L. Barab\'asi, F. Caserta, S. Havlin, H. E. Stanley,
and T. Vicsek, Phys. Rev. A {\bf 45} (1992) R8313.

\item{[21]} A.B.Harris, J.Phys.C7 (1974) 1671.

\item{[22]} For comparison: the EW eq. has $\zeta = 1/2, ~\beta =1/4 $
in $D=1$, while for $D=2$ the roughness diverges only logarithmically [14];
the KPZ exponents are $\zeta =1/2, ~\beta=1/3 $ in $D=1$ and $\zeta \simeq
0.39, \beta \simeq 0.24 $ in 2+1 dimensions [1].

\bigskip
\bigskip
\bigskip

\noindent{\bf Note added:}

\noindent We determined the velocity exponent $\theta $ in 1+1
dimemnsions by a
simululation of large systems (up to $L=262144$) to $\theta = 0.25 \pm 0.03$.
(This took 30 h CPU on one Cray Y-MP processor.)

\vfil\eject

{\bf Figure captions}

\item{Fig.1} The width $w^2 (L)$ for $g=1,~p=p_c\simeq 0.8004$ (squares)
and for $g=2, ~p=p_c \simeq 0.8748$ (circles).
The statistical
uncertainities are smaller than the size of the symbols.

\item{Fig.2} The effective roughness exponent $\zeta(L)$ for $g=1,~p=p_c$
(squares) and $g=2,~p=0.8744<p_c$ (circles). For $p<p_c$ an interface
is pinned at earlier times and
one is able to simulate larger systems than for $p=p_c$.

\item{Fig.3} Scaling of $H(t)$ for $L=262144$ and $p=p_c=0.8004$ ($g=1$).

\item{Fig.4} The width $w^2(t)$ for the same runs as in fig.3.

\item{Fig.5} The effective exponents $\beta (t)$ determined from the height
$H(t)$ (circles) and from the width $w^2(t)$ (squares).
In fig. 5b $ \beta (t)$ is determined from a plot $w^2 (2t) - w^2 (t) $
(and analogous $H(2t) -H(t)$), whereas in fig 5a the exponents are
calculated from a simple plot $w^2 (t)$ and $H(t)$.

\item{Fig.6} Best scaling plot of the height-height correlation function
$C(r,t)$, which is scaled by $t^\beta$, $\beta =0.88$ versus
$r/t^{\beta/\zeta}$,
$\zeta =1.25$. The same plotting symbol is used for data at a given time $t$.
In fig. 6a all data are plotted and in fig. 6b the
crossover region is shown only.

\item{Fig.7} The effective exponents $\zeta (L) $ for $g=4,~p=p_c\simeq 0.6416$
(squares) and $g=6,~p=p_c \simeq 0.74448$ (circles).

\item{Fig.8} The interface width $w^2(t)$ for
$L^2 =1024^2, ~g=4,~p=p_c=0.6416$.

\item{Fig.9} The effective exponents $\beta (t)$ determined from
the plot of $w^2(t)$
(squares) and from $H(t)$ (circles) for $g=4,~p=p_c=0.6416$.
In addition, $\beta (t)$ from the height
for the case $p=0.643>p_c$ (triangles) is plotted for comparison
(see text).

\item{Fig.10} Double logarithmic plot of the velocity versus $p-p_c$
for $g=4$ (squares) and $g=6$ (circles).

\vfil\eject
\bye